\documentclass[lettersize,journal]{IEEEtran} 
\usepackage{cite}
\usepackage{booktabs}
\usepackage{mathrsfs}
\usepackage{etoolbox}
\makeatletter
\patchcmd{\IEEEbiographynophoto}
  {\vskip 4\baselineskip plus 1fil minus 0\baselineskip}
  {\vspace{0.6\baselineskip}}
  {}{}
\makeatother

\usepackage[colorlinks]{hyperref}
\usepackage{amsthm}
\usepackage{caption}
\DeclareCaptionFormat{myformat}{\textbf{#1#2} #3}
\usepackage[cmex10]{amsmath}
\usepackage{mathtools}
 \usepackage{siunitx}
\usepackage{array}
\usepackage{color}
\usepackage{tabularray}
\usepackage{eqparbox}
\usepackage{bm}

\usepackage[table,dvipsnames]{xcolor}
\usepackage{multicol,booktabs,tabularx}

\usepackage{cases}
\usepackage{algorithm}
\usepackage{paralist}
\usepackage{algpseudocode}

\usepackage{graphicx}
\usepackage{subfigure}
\usepackage{epstopdf}
\usepackage{epsfig}
\usepackage{amssymb}
\usepackage{array}
\usepackage{multirow}

\title{AI/ML for Mobile Networks: Release 19 Status and Future Challenges}
\author{Yuan Gao, \textit{Member, IEEE},
Xinyi Wu, Jun Jiang, Bintao Hu, Jianbo Du, Qiang (John) Ye, \textit{Senior Member, IEEE},\\
Shunqing Zhang, \textit{Senior Member, IEEE},
F. Richard Yu, \textit{Fellow, IEEE}, Shugong Xu, \textit{Fellow, IEEE}

\thanks{This work was supported by Shanghai Natural Science Foundation under Grant 25ZR1402148. (Shugong Xu is the corresponding author)}
\thanks{Yuan Gao, Xinyi Wu and Shunqing Zhang are with the School of Communication and Information Engineering, Shanghai University, China, email: gaoyuansie@shu.edu.cn, wu\_xinyi0312@shu.edu.cn and shunqing@shu.edu.cn.}
\thanks{Jun Jiang and Shugong Xu are with Xi’an Jiaotong-Liverpool University, Suzhou, China, email: jun.jiang25@student.xjtlu.edu.cn, shugong.xu@xjtlu.edu.cn.}
\thanks{Bintao Hu is with School of IoT, Xi’an Jiaotong-Liverpool University, Suzhou, China, email: Bintao.Hu@xjtlu.edu.cn.}
\thanks{Jianbo Du is with the School of Communication and Information Engineering, Xi'an University of Posts and Telecommunications, Xi'an, China, e-mail: dujianboo@163.com.}
\thanks{Qiang (John) Ye is with the Department of Electrical and Software Engineering, Schulich School of Engineering, University of Calgary, Canada, e-mail: qiang.ye@ucalgary.ca.}
\thanks{F. Richard Yu is with the School of Information Technology, Carleton University, Ottawa, Canada. E-mail: richard.yu@carleton.ca.}
}

\begin{document}
\maketitle
\begin{abstract}

Artificial intelligence (AI) and machine learning (ML) have become essential drivers of sixth-generation (6G) mobile networks. Unlike earlier surveys that mainly review high-level standardization progress, this paper focuses on the practical deployment barriers that remain after Release 19. We identify three open problems: dataset construction under 3GPP guidelines, cross-scenario and cross-simulator generalization, and baseline selection under joint performance-cost KPIs. These problems are examined through a CSI-feedback case study that uses both DeepMIMO (outdoor/indoor, ray-tracing) and QuaDRiGa (indoor-factory/highway, geometry-based stochastic) channel models. It reveals pronounced generalization challenges in zero-shot transfer, especially when models move across different simulators. A clear mismatch exists between deterministic ray-tracing and stochastic geometry-based modeling paradigms and leads to dramatic performance degradation. In contrast, pre-training on a source simulator followed by targeted fine-tuning on the target domain substantially mitigates these cross-simulator difficulties, and markedly outperforms zero-shot transfer. Attention-based models generalize better across the two physics regimes while keeping computational cost moderate. These results explain why the fine-tuning and LCM procedures already partially standardized in Release 19 are necessary, and they point to concrete next steps for scalable LCM, wider use cases, and multi-task learning.

\end{abstract}
\section{Introduction}

As mobile networks continue to evolve, the upcoming sixth-generation (6G) systems bring unprecedented levels of complexity. These advanced networks aim to deliver ubiquitous connectivity across terrestrial, aerial, and space domains. They will operate over an extremely wide frequency range, from sub-\SI{6}{\giga\hertz} to \SI{}{\tera\hertz}\cite{gao2026csiextra}. Achieving this vision requires the integration of sophisticated technologies such as extremely large-scale multiple-input multiple-output (MIMO) and high-capacity access schemes. However, managing such intricate systems far exceeds the capabilities of conventional algorithms. This situation creates a compelling opportunity for artificial intelligence (AI) and machine learning (ML) to address these emerging challenges.

Recognizing the strong potential of AI/ML, the 3rd Generation Partnership Project (3GPP) has actively worked to standardize their integration into mobile networks. Release 17 established high-level principles for AI-enabled radio access network (RAN) intelligence \cite{37_817}. Release 18 focused on improving data collection mechanisms and introducing signaling support for AI/ML-based networks \cite{chen20235g}. To systematically evaluate the benefits and costs of AI applications in mobile communications, 3GPP Technical Report 38.843 defines three representative use cases, channel state information (CSI) feedback, beam management, and positioning enhancement, along with their corresponding key performance indicators (KPIs) and data generation guidelines \cite{38_843}.

Standardization efforts have progressed rapidly through Release 19 and beyond. These developments have introduced normative specifications and deeper insights across both the air interface and the broader 6G system. Several recent studies have summarized these advancements, with their main focuses outlined in Table \ref{tab:ai_ml_papers}. Most of them concentrate on high-level standardization progress in Release 18 \cite{lin2023artificial,shehzad2022artificial,yeh2024standardization,jin2026generalizable}. One recent work reviews Release 19 enhancements from the service and architecture perspective, covering both `Network for AI' (model distribution, federated learning, D2D support) and `AI for Network' (network data analytics function (NWDAF) improvements for location, policy, and signaling) aspects \cite{taksande2025ai}. Nevertheless, these reviews generally provide limited discussion on practical deployment challenges. Important topics such as dataset quality and preparation, cross-scenario and cross-dataset generalization, baseline model comparisons, computational complexity, inference latency, and performance–cost trade-offs remain largely underexplored.

\begin{table*}[htbp]
\centering
\caption{Main focuses of the existing research on AI/ML standardization}
\label{tab:ai_ml_papers}
\begin{tabular}{l p{15cm} }
\toprule
 \textbf{Reference} & \textbf{Main focus to AI/ML Standardization}  \\
\midrule
\cite{shehzad2022artificial}, 2022 
& The article reviews ML paradigms and case studies (e.g., UE positioning) for PHY and link-layer applications while outlining standardization timelines and industrial gaps for 6G. 
\\
\midrule

\cite{lin2023artificial}, 2023 
& A holistic examination of AI/ML activities across all 3GPP Rel 18 working groups, explaining design rationales for model transfer, system support, security, media, and NR air interface as a bridge to 6G. 
\\
\midrule
\cite{jin2026generalizable}, 2024 
& This work provides a standardization-focused comparison of legacy and AI-enabled beam management frameworks in 6G, detailing spatial and temporal beam prediction procedures to reduce overhead and latency. 
\\
\midrule

\cite{yeh2024standardization}, 2024 
& The integration of 3GPP standardization efforts in O-RAN RIC for emerging trends such as federated learning, E2E learning, and XAI. 
 \\
\midrule

\cite{taksande2025ai}, 2025 
& Advancement in Rel 19 from the perspective of service aspect, including core network, media services, application-layer enablement framework. 
 \\\midrule
 This paper &
Focuses on the practical deployment barriers of AI/ML for the NR air interface by quantifying cross-simulator domain shift in CSI feedback and showing why the fine-tuning and LCM mechanisms of Release 19 are necessary.\\
\bottomrule
\end{tabular}
\vspace{-6mm}
\end{table*}

To address these gaps, this paper focuses on the practical deployment barriers that remain after Release 19. The contributions are:
\begin{enumerate}
    \item We analyze the 3GPP AI/ML framework up to Release 19 from a deployment viewpoint, focusing on three open problems: dataset construction under the parameter settings in TR 38.843, generalization evaluation, and baseline selection under joint performance-cost KPIs.
    \item We evaluate the cross-simulator domain shift with a CSI-feedback case study that uses DeepMIMO (outdoor/indoor, ray-tracing) and QuaDRiGa (indoor-factory/highway, geometry-based stochastic) channel models. Zero-shot transfer fails across the two modeling paradigms. Source-domain pre-training followed by target-domain fine-tuning recovers most of the performance loss.
    \item The case study demonstrates that compact attention architectures capture environment-agnostic structural patterns more effectively than classical CNN or MLP designs under the evaluated metrics, including NMSE, FLOPs, model size, latency, and peak memory, while offering a favorable accuracy–memory trade-off for UE-side models.
    \item From these results we identify research directions that address the remaining barriers: scalable LCM, expansion beyond current air-interface use cases, and multi-task learning for AI-native 6G networks.
\end{enumerate}


\section{A Holistic Picture of AI/ML in 3GPP}
\begin{figure*}
  \centering
\includegraphics[width=0.86\textwidth]{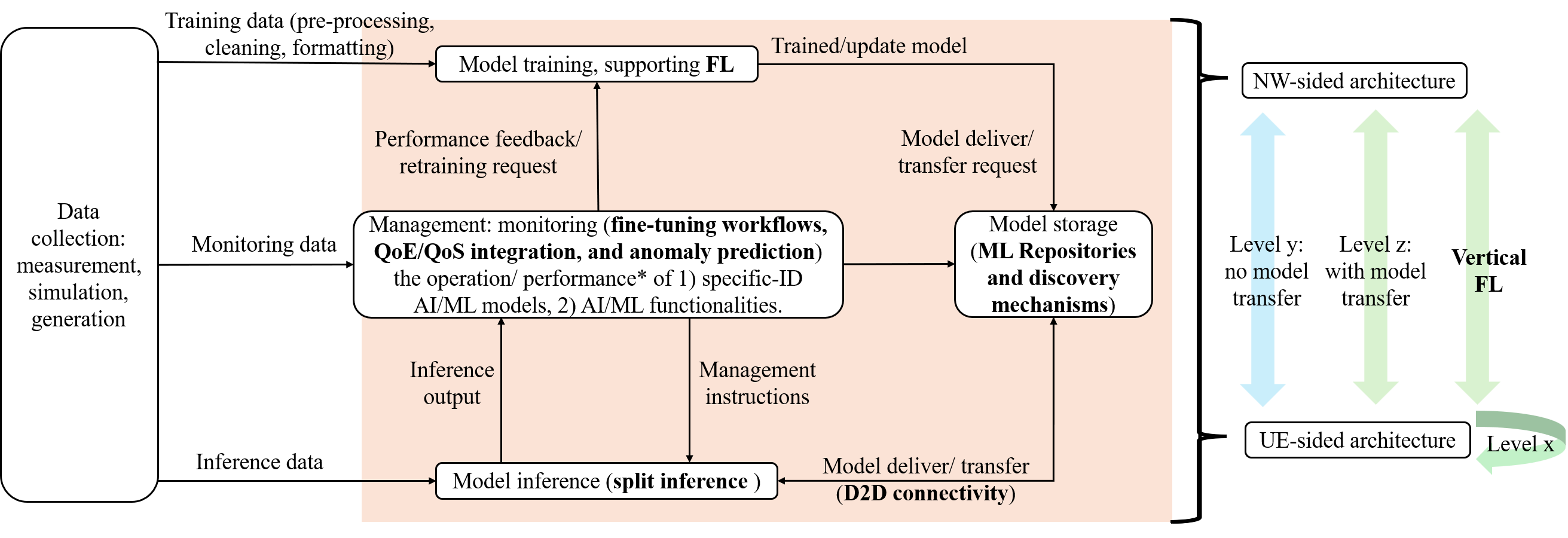}
 \caption{The architecture of LCM of AI/ML models at UE and NW sides, and the collaboration between UE and NW side models. The advancements in Release 19 and beyond are in \textbf{bold}.}
  \label{architecture}
  \vspace{-3mm}
\end{figure*}
\begin{table*}[]
\caption{Evaluations of AI/ML for representative use cases defined in 3GPP\cite{38_843}.}
\label{Evaluations}
\begin{tabular}{|c|c|c|cc|}
\hline
\multirow{2}{*}{Use case} &
  \multirow{2}{*}{Sub use case} &
  \multirow{2}{*}{Use-case-specific link-level performance metrics} &
  \multicolumn{2}{c|}{Common KPIs} \\ \cline{4-5} 
 &
   &
   &
  \multicolumn{1}{c|}{Performance} &
  Cost \\ \hline
\multirow{2}{*}{\begin{tabular}[c]{@{}c@{}}CSI\\feedback\end{tabular}} &
  CSI compression &
  SGCS/NMSE of output CSI &
  \multicolumn{1}{c|}{\multirow{6}{*}{\begin{tabular}[c]{@{}c@{}}System level performance\\ e.g., throughput\\ (mandatory),  generalization \\performance\end{tabular}}} &
  \multirow{6}{*}{\begin{tabular}[c]{@{}c@{}}Over-the-air overhead,\\ inference complexity, \\ training complexity, \\ LCM-related complexity,\\ storage requirements\end{tabular}} \\ \cline{2-3}
 &
  CSI prediction &
  SGCS/NMSE of each predicted instance &
  \multicolumn{1}{c|}{} &
   \\ \cline{1-3}
\multirow{2}{*}{BM} &
  Spatial-domain DL beam prediction &
  \multirow{2}{*}{\begin{tabular}[c]{@{}c@{}}Beam prediction accuracy, reduction of RS\\ overhead, UCI, latency, power consumption\end{tabular}} &
  \multicolumn{1}{c|}{} &
   \\ \cline{2-2}
 &
  Time-domain DL beam prediction &
   &
  \multicolumn{1}{c|}{} &
   \\ \cline{1-3}
\multirow{2}{*}{Positioning} &
  Direct positioning &
  90\% percentile of horizontal accuracy &
  \multicolumn{1}{c|}{} &
   \\ \cline{2-3}
 &
  AI-assisted positioning &
  Ranging and angle estimation error &
  \multicolumn{1}{c|}{} &
   \\ \hline
\end{tabular}
\vspace{-6mm}
\end{table*}

\subsection{General AI/ML framework}
\subsubsection{LCM of AI Models}
In 3GPP Release~18, the foundational framework for AI and ML model LCM was established. The work focused primarily on the NR air interface and adopted a radio access network (RAN)-centric approach. Key modules include model identification through two complementary methods. Functionality-based LCM integrates with existing user equipment (UE) capability signaling. Model ID-based LCM relies on unique model identifiers for mutual recognition between the network and the UE. Standardized lifecycle operations, such as selection, activation, deactivation, switching, and fallback, were defined via signaling procedures.

Model training was specified with three distinct collaboration types to ensure multi-vendor interoperability. Type~1 performs joint training on one side only. Type~2 conducts concurrent joint training across both sides. Type~3 separates training while sharing datasets. Release~19 primarily addresses one-sided models at Level~y \cite{38_843}. Further normative work on two-sided models is expected in subsequent releases\footnote{Qualcomm, `New WI: Artificial Intelligence (AI)/Machine Learning (ML) for NR air interface enhancements,' 3GPP, RP-251870, V15.1.1
(2025-06).}\textsuperscript{, }\footnote{Nokia (moderator), `New SI: Study on Artificial Intelligence (AI)/Machine Learning (ML) for NG-RAN Phase 3,' 3GPP, RP-251868, V15.3.1
(2025-06).}. Additional mechanisms address model distribution and transfer, model update and fine-tuning, generalization, and performance monitoring. These include reliable deployment strategies, mandatory performance-metric reporting, and rapid fallback to legacy baselines when needed. Initial support for federated learning (FL) and advanced collaboration was outlined through three levels of network-UE interaction.

Release~19 extended LCM support beyond the earlier RAN-focused study. It introduced normative specifications for a generic lifecycle-management framework of one-sided models, with further work continuing under the Service and System Aspects (SA) working group. Enhancements span multiple modules. Refined model identification, storage, and management now include ML repositories and discovery mechanisms in the application layer and operations, administration and maintenance/management data analytics function (OAM/MDAF). Lifecycle operations and model training were expanded to support dynamic FL member addition or removal, vertical FL (VFL), transfer learning, reinforcement learning, and distributed training coordination via NWDAF model training logical function (MTLF) and MDAF. Model distribution gained robust device-to-device connectivity support for split inference and distributed learning. Model update, generalization, and performance monitoring were strengthened with unified OAM-level fine-tuning workflows, quality of experience/quality of service (QoE/QoS) integration, and anomaly prediction. FL and advanced collaboration mechanisms were further developed at the system level, including application-layer enablement and support for hierarchical deployment. These changes aim to support more practical AI/ML deployment across the core network, media services, and vertical applications.

\subsection{Use Cases}

To evaluate the benefits and costs of AI/ML in mobile networks, 3GPP identified three key use cases at the air-interface level: CSI feedback, beam management, and positioning.

Acquiring accurate CSI lies at the core of mobile networks. Release~18 studied AI/ML-based CSI feedback for both spatial-frequency domain compression and time-domain prediction. The former reconstructs CSI in frequency-division duplex (FDD) systems and reduces uplink overhead. The latter mitigates channel aging caused by the delay between CSI reporting and actual use at the gNB. Release 19 moved time-domain CSI prediction with one-sided UE models into normative work \cite{38_843}. Two-sided CSI compression remains under further study and is expected to be addressed in subsequent releases \cite{huang2025ai}. Key aspects include model pairing, LCM signaling, and uplink control information (UCI) mapping\footnote{Qualcomm (moderator), `Revised WID: Artificial Intelligence (AI)/Machine Learning (ML) for NR air interface Phase 2,' 3GPP, RP-260511 (2025-03).}.

Effective beam management is essential for interference mitigation, coverage extension, and capacity enhancement, particularly in millimeter-wave networks. 3GPP defined two sub-use cases. Spatial-domain downlink beam prediction exploits feedback from one set of downlink beams to forecast the optimal beam from another set. Time-domain downlink beam prediction uses historical measurements from the same beam set to predict future beams \cite{jin2026generalizable}. Release~19 introduced normative specifications for both spatial-domain and time-domain downlink beam prediction using one-sided models \cite{38_843}. The specifications define the necessary signaling so that gNB- or UE-sided models can be used while maintaining backward compatibility with existing frameworks. Future work will focus on enhancing these capabilities with more complex scenarios such as multiple transmission reception point (multi-TRP) and mobility optimization, along with deeper LCM integration for practical deployment\footnote{Samsung (moderator), `Revised WID: Topic Summary for [117][107] 6G AI,' 3GPP, R4-2522272 (2025-11).}.

Positioning forms the foundation for many sensing functions \cite{xu2025enhanced}. 3GPP defined two sub-use cases. Direct AI/ML positioning determines UE location, for example through fingerprinting based on CSI. AI/ML-assisted positioning generates new estimates or refines existing ones, such as identifying line-of-sight (LOS)/non-line-of-sight (NLOS) conditions or improving ranging and angle measurements. Release~19 advanced both direct and AI-assisted positioning into normative work. Models can be placed at the UE, gNB, or LMF. Enhanced measurements such as LOS/NLOS probability are included to improve accuracy in NLOS and indoor scenarios. Further integration with NWDAF and LCM mechanisms is under development \cite{38_273}.

\subsection{Performance Evaluations}
3GPP has established a comprehensive set of performance metrics. These metrics assess both the benefits and the costs of integrating AI/ML models into mobile networks. As demonstrated in Table \ref{Evaluations}, they consist of common KPIs that apply across all use cases and use-case-specific link-level indicators.

Common performance KPIs emphasize system-level gains, such as throughput and generalization capability across scenarios. Cost-related KPIs address over-the-air signaling overhead, inference complexity, training complexity, LCM-related overhead, and storage requirements. These cover radio resources used for AI-related signaling, model size, computational demands for inference and training, and overhead from model management operations, such as activation, deactivation, switching, and fallback.

Use-case-specific metrics evaluate direct impact on physical-layer operations. For CSI feedback, squared generalized cosine similarity (SGCS) and normalized mean squared error (NMSE) measure reconstruction accuracy in compression tasks. Per-instance metrics are used for prediction. For beam management, key indicators include beam prediction accuracy, reduction in reference-signal overhead, UCI overhead, latency, power consumption, and system-level throughput statistics. For positioning, the primary metric is the 90th-percentile horizontal accuracy, while ranging and angle estimation errors serve as intermediate indicators in AI/ML-assisted schemes.

\section{Key Research Problems: \\Opportunities and Challenges}
There is broad consensus on the potential benefits of AI/ML for mobile network management and operations. However, realizing this vision still faces significant challenges. These challenges mainly arise from three critical areas: dataset preparation, generalization evaluation, and baseline model selection.

\subsection{Dataset Preparation }

Datasets play a central role in the performance of AI/ML models. This holds true for mobile networks as well. Nevertheless, no high-quality common dataset currently exists for evaluating AI/ML models in mobile networks, particularly for the representative use cases discussed earlier.
\subsubsection{Parameter settings of 3GPP}
3GPP does not provide any standard dataset for the representative use cases. Instead, it specifies detailed parameter settings for dataset construction during the training, validation, and testing phases~\cite{38_843}. Given the practical difficulties in acquiring high-quality real-world measurement data, research institutes and companies are encouraged to generate datasets through simulation. The essential requirement is that all parameters remain fully aligned with the defined guidelines.

To standardize dataset generation, 3GPP has established both common parameter settings and use-case-specific parameter settings. The common parameters fall into two main categories. The first category addresses deployment scenarios (for example, urban macro (UMa), urban micro (UMi), and indoor hotspot (InH)). It also includes outdoor or indoor UE distributions, carrier frequencies, and other configurable elements such as antenna spacings, inter-site distances, and UE speeds. The second category covers system configurations, including bandwidths, antenna port layouts, numerologies, and the number of transmission ranks or layers.

Additional parameters are recommended for each use case. For CSI feedback, CSI feedback payloads must be considered during data generation. For spatial-domain downlink (DL) beam prediction, downlink transmit (Tx) beam codebook configurations are critical. For time-domain beam prediction, the UE trajectory model plays an important role. For positioning, environmental factors such as clutter parameters and time-varying changes (e.g., mobility of clutter objects) are required. Imperfect factors, including network synchronization error (e.g., training dataset without synchronization error versus test dataset with it) and UE/gNB RX and TX timing error, must also be included.

\textit{Remarks}: Common parameter configurations provide a broad framework for the wireless environment. In contrast, use-case-specific parameters are more detailed and intricate. They vary significantly depending on the particular application scenario.
\subsubsection{Discussions}
Current datasets used for evaluating AI/ML in wireless communications can be classified into simulated and measured datasets. Their features and limitations are detailed below.

Simulated datasets can be further divided into ray-tracing (RT)-based and stochastic channel model-based datasets, including those using geometry-based stochastic channel models (GSCM). The major limitation of RT-based datasets is that they cannot model time-varying changes. Examples include DeepMIMO-based\footnote{https://github.com/DeepMIMO/DeepMIMO-matlab\label{deepMIMO}},  WAIR-D\footnote{https://www.mobileai-dataset.com/html/default/yingwen/DateSet/15909942\\53188792322.html?index=1}, etc. In addition, they do not contain the additional parameters needed for positioning evaluation, such as synchronization error and RX/TX timing error. GSCM-based datasets, such as xG-Loc\footnote{https://ieee-dataport.org/open-access/xg-loc-3gpp-compliant-datasets-xg-location-aware-networks} and wireless intelligence\footnote{https://wireless-intelligence.com/\#/dataSet?id=2c92185c7e3f1aa4017e3f2\\b9d6e0000
}, are inherently statistical. They may still simplify certain hardware or environmental non-idealities, resulting in a persistent simulation-to-reality gap.

Measured datasets also have their own limitations due to the specific measurement setups. These include the absence of outdoor data (e.g., Time-varying industrial CSI dataset\footnote{https://ieee-dataport.org/documents/data-set-time-varying-industrial-radio-channels-varying-environment}), the absence of indoor data (DeepSense 6G\footnote{https://www.deepsense6g.net/}), and the absence of frequency range 2 (FR2) data (DICHASUS\footnote{https://dichasus.inue.uni-stuttgart.de/}).


\subsection{Generalization Evaluation}
Generalization performance is a significant indicator of the effectiveness of AI/ML models across various scenarios. 3GPP Technical Report 38.843 emphasizes the importance of generalization performance~\cite{38_843}. The generalization tests defined by 3GPP are as follows:
\begin{inparaenum}[\itshape a\upshape)]
\item Case 1: the training and test sets are disjoint sets from the dataset A\footnote{Dataset A and B can be generated using a single scenario/configurations or a mix a multiple scenario/configurations as indicated in the section of Dataset}; 
\item Case 2: the training and test sets are from dataset A and B, respectively;
\item Case 3: the model is trained using dataset A and is further fine-tuned and tested using a disjoint set from dataset B; and
\item Case 4: the training is a mix of dataset A and a subset from dataset B, and the test set is a disjoint set from dataset B.
\end{inparaenum}

Current research on AI models for mobile networks predominantly uses a single dataset for both training and testing (Case 1). This approach falls short in developing models with the robust generalization capabilities required for real-world applications. Case 2 demands the highest level of generalization ability. Case 3 follows next, while Case 4 aims to optimize performance across multiple datasets simultaneously.

\textit{Remarks}: Existing studies mainly focus on Case 1. To enhance the generalization capabilities of AI models, future research should prioritize training strategies that incorporate Cases 2, 3, and 4.


\subsection{Selection of AI/ML model}
Selecting appropriate baseline AI/ML models, whether for general use or specific use cases, is crucial for future research and performance benchmarking. However, 3GPP has not recommended any baseline AI models~\cite{38_843}. Furthermore, evaluations of various AI models have yielded conflicting results, as evidenced in channel prediction research~\cite{stenhammar2024comparison}. In addition, as indicated in Table~\ref{Evaluations}, multiple KPIs must be considered jointly to evaluate AI/ML models. This requirement makes the choice of AI models even more challenging. Consequently, the optimal selection of AI models for mobile communications remains an open question.

\begin{figure} \centering 
  \centering
\includegraphics[width=0.36\textwidth]{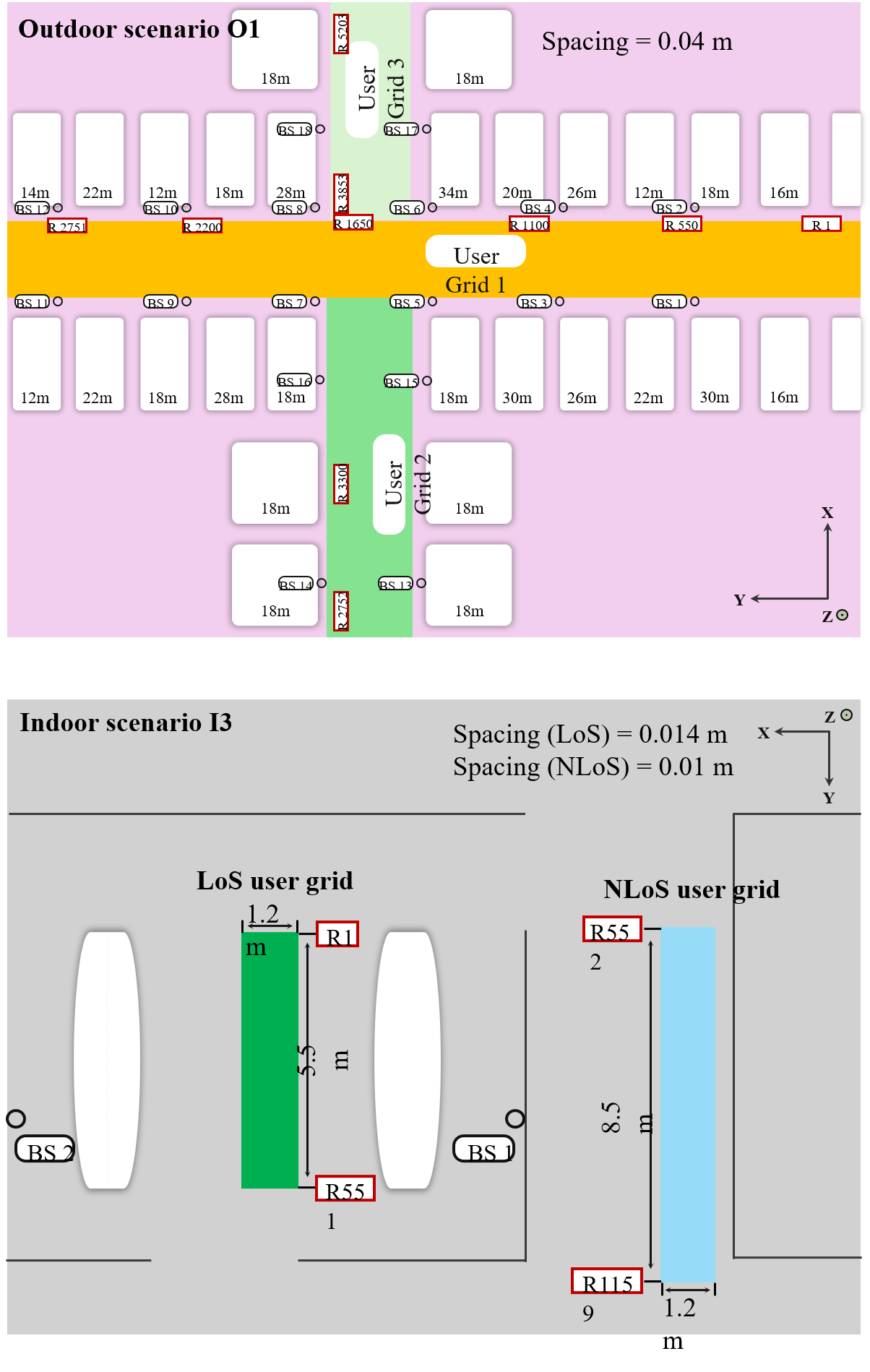}
 \caption{Outdoor and indoor scenarios to generate CSI data using DeepMIMO.}
  \label{architecture}     
  \vspace{-6mm}
\end{figure}

\begin{table*}[!t]
\caption{Simulation results of CSI feedback with compression ratio of $1/4$ in terms of cost and performance in NMSE (dB) in different training scenarios. The cost of models are evaluated in terms of floating-point operations (FLOPs) (M), model size (M), end-2-end (E2E) latency (ms) and peak memory (M) using  RTX 3090. `In', `Out', `InF' and `Hwy' denote indoor, outdoor, indoor-factory and highway scenarios, respectively. `*-Z' and `*-F' denote zero-shot learning and full-model fine-tuning in the target scenario `*'. Best performance in each column are in \textbf{bold}.}
\label{tab:cross_domain_full}
\centering
\begin{tabular}{cc|cccccccc|cccc}
\hline
\begin{tabular}[c]{@{}c@{}}Trained \\ scenario\end{tabular} &
  Model &
  In-Z &
  Out-Z &
  InF-Z &
  Hwy-Z &
  In-F &
  Out-F &
  InF-F &
  Hwy-F &
  FLOPs &
  Model size &
 Latency &
  Peak memory \\ \hline
\multirow{5}{*}{In} &
  CsiNet &
  -20.1 &
  -8.8 &
  -5.4 &
  -9.3 &
  - &
  -20.8 &
  -17.3 &
  -24.0 &
  5.6 &
  \textbf{8.0} &
  \textbf{0.9} &
 \textbf{ 17.2} \\
 &
  MNet &
  -28.5 &
  \textbf{-8.9} &
  \textbf{-6.1} &
  \textbf{-10.1} &
  - &
  -24.1 &
  -25 &
  -32.7 &
  \textbf{4.0} &
  8.2 &
  1.4 &
  17.4 \\
 &
  DCRNet &
  -25.2 &
  -8 &
  -4.6 &
  -8.9 &
  - &
  -25.6 &
  -21.8 &
  -25.1 &
  4.4 &
  8.0 &
  3.3 &
  17.3 \\
 &
  TransNet &
  -25.4 &
  -8.1 &
  -2.6 &
  -9.3 &
  - &
  -24.7 &
  -22.5 &
  -25.2 &
  35.7 &
  10.2 &
  2.1 &
  19.7 \\
 &
  StarCANet &
  \textbf{-35.2} &
  -8.8 &
  -5.7 &
  -9.4 &
  - &
  \textbf{-35.5} &
  \textbf{-30.3} &
  \textbf{-38.3} &
  20.6 &
  8.1 &
  4.4 &
  17.3 \\ \hline
\multirow{5}{*}{Out} &
  CsiNet &
  -9.9 &
  -20 &
  -5.5 &
  -9.3 &
  -23.9 &
  - &
  -17.5 &
  -24.0 &
  5.6 &
  \textbf{8.0} &
  \textbf{0.9} &
  \textbf{17.2} \\
 &
  MNet &
  -10.7 &
  -23.6 &
  -5.2 &
  \textbf{-10.1} &
  -29 &
  - &
  -24.9 &
  -32.7 &
  \textbf{4.0} &
  8.2 &
  1.4 &
  17.4 \\
 &
  DCRNet &
  -9.6 &
  -22.8 &
  -4.1 &
  -8.9 &
  -28.3 &
  - &
  -27.3 &
  -25.1 &
  4.4 &
  8.0 &
  3.3 &
  17.3 \\
 &
  TransNet &
  -10.3 &
  -24.1 &
  -4 &
  -9.3 &
  -26.7 &
  - &
  -25.7 &
  -25.2 &
  35.7 &
  10.2 &
  2.1 &
  19.7 \\
 &
  StarCANet &
  \textbf{-11.2} &
  \textbf{-34.9} &
  \textbf{-5.6} &
  -9.4 &
  \textbf{-41.2} &
  - &
  \textbf{-31.5} &
  \textbf{-38.3} &
  20.6 &
  8.1 &
  4.4 &
  17.3 \\ \hline
\multirow{5}{*}{InF} &
  CsiNet &
  -14.0 &
  -7.9 &
  -15.7 &
  -22.2 &
  -18.3 &
  -13.4 &
  - &
  -22.1 &
  5.6 &
  \textbf{8.0} &
  \textbf{0.9} &
  \textbf{17.2} \\
 &
  MNet &
  -22.8 &
  -8.6 &
  -23.1 &
  -31.1 &
  -23.7 &
  \textbf{-24.0} &
  - &
  -24.4 &
  \textbf{4.0} &
  8.2 &
  1.4 &
  17.4 \\
 &
  DCRNet &
  -24.8 &
  -8.7 &
  -24.1 &
  -28.1 &
  -26.0 &
  -21.3 &
  - &
  -22.9 &
  4.4 &
  8.0 &
  3.3 &
  17.3 \\
 &
  TransNet &
  -16.4 &
  -7.6 &
  -23.9 &
  -11.4 &
  -26.0 &
  -22.3 &
  - &
  -25.3 &
  35.7 &
  10.2 &
  2.1 &
  19.7 \\
 &
  StarCANet &
  \textbf{-26.1} &
  \textbf{-9.3} &
  \textbf{-30.6} &
  \textbf{-38.5} &
  -\textbf{28.3} &
  -23.3 &
  - &
  -\textbf{27.6} &
  20.6 &
  8.1 &
  4.4 &
  17.3 \\ \hline
\multirow{5}{*}{Hwy} &
  CsiNet &
  \textbf{-10.7} &
  -8.7 &
  -8 &
  -21.7 &
  -14.6 &
  -10.5 &
  -10.2 &
  - &
  5.6 &
  \textbf{8.0} &
  \textbf{0.9} &
  \textbf{17.2} \\
 &
  MNet &
  -10 &
  -7.9 &
  -6.8 &
  -31.6 &
  -\textbf{28.5} &
  -22.3 &
  -22.4 &
  - &
  \textbf{4.0} &
  8.2 &
  1.4 &
  17.4 \\
 &
  DCRNet &
  -8.3 &
  -8.4 &
  -7.2 &
  -25.7 &
  -21.6 &
  -19.7 &
  -20.3 &
  - &
  4.4 &
  8.0 &
  3.3 &
  17.3 \\
 &
  TransNet &
  -10.2 &
  -6.5 &
  -9 &
  -24.4 &
  -25.6 &
  -21.0 &
  -22.1 &
  - &
  35.7 &
  10.2 &
  2.1 &
  19.7 \\
 &
  StarCANet &
  -10.6 &
  \textbf{-9.2} &
  \textbf{-10.1} &
  \textbf{-33.9} &
  -26.6 &
  -\textbf{29.0} &
  -\textbf{26.8} &
  - &
  20.6 &
  8.1 &
  4.4 &
  17.3 \\ \hline
\end{tabular}
\vspace{-6mm}

\end{table*}

\section{Case study}
This section presents a detailed case study on CSI feedback. We compare several representative AI/ML models in terms of reconstruction performance, generalization capability, and computational complexity. The primary goal is to move beyond commonly recognized challenges such as dataset scarcity and domain shifts. The study therefore delivers practical, scenario-specific insights into real-world deployment barriers as well as the architectural strengths of AI-based CSI feedback solutions in realistic 6G environments.

\subsection{Simulation settings and CSI generation}
In the case study, we employed four scenarios that represent different channel modeling paradigms and typical 6G use cases.

We generated four datasets in total. Two were created with the DeepMIMO simulator: an outdoor urban scenario (O1) at 3.5 GHz and an indoor scenario (I3) at 2.4 GHz. As illustrated in Fig.~\ref{architecture}, O1 consists of two streets with an intersection and I3 represents a $10~\mathrm{m} \times 11~\mathrm{m} \times 3~\mathrm{m}$ indoor conference room containing eight human models and two conference tables. After filtering out users without valid propagation paths, O1 contains 62,910 valid CSI samples and I3 contains 65,493 samples. The DeepMIMO datasets were split into training and validation sets at a 4:1 ratio. More detailed configurations of O1 and I3 can be found on the DeepMIMO homepage\textsuperscript{\ref{deepMIMO}}

The remaining two datasets were generated using the QuaDRiGa simulator at 2.4 GHz to represent industrial and high-mobility conditions. These include the Indoor Factory (InF) scenario, which features rich multipath propagation due to metallic structures and machinery, and the Highway scenario, which involves high-speed mobility, Doppler effects, and dynamic LOS/NLOS transitions. In all scenarios, the base station is equipped with a 32-antenna uniform linear array (ULA), while each UE has a single antenna. The system bandwidth is 100 MHz with 256 subcarriers. In the Highway scenario, each UE generates 10 frames at 10 ms intervals, and the UE speed is set to 140 km/h. For both InF and Highway scenarios, 50,000 samples were used for training and 5,000 samples for testing.

For a fair comparison, we evaluated five representative CSI feedback baseline models that cover a range of architectural paradigms. CsiNet~\cite{chakma2023deep} is a classical convolutional neural network (CNN)-based CSI compression model and serves as a widely used benchmark. MNet is a lightweight multilayer perceptron (MLP)-based model designed for low-complexity feedback. DCRNet enhances CNN feature extraction through residual connections and dilated convolutions. TransNet employs self-attention mechanisms for CSI reconstruction. Finally, StarCANet~\cite{zhao2025starcanet} represents recent compact attention-based designs. All baseline models were trained using the same data split. The batch size was set to 512, and each model was trained for 1000 epochs with a fixed random seed. We adopted the mean squared error (MSE) loss function and the Adam optimizer with an initial learning rate of $  1 \times 10^{-3}  $. The learning rate was adaptively adjusted using the ReduceLROnPlateau scheduler with a decay factor of 0.8. All experiments were performed on an NVIDIA GeForce RTX 3090 graphics processing unit (GPU). For fine-tuning, each model was initialized from its best source-domain checkpoint, instead of random initialization, and trained on the complete target-domain training split. All encoder and decoder parameters remained trainable. All other settings, including the number of epochs, batch size, optimizer, learning-rate schedule, and random seed, were identical to those used for initial training.

These models were selected because they represent the major evolutionary stages of CSI feedback architectures. They range from early CNN-based designs to residual and dilated CNNs, MLP-based approaches, transformer-based models, and modern efficient attention mechanisms. Moreover, they span a wide range of computational complexity, with floating-point operations (FLOPs) ranging from 2.0 million to 35.7 million. This representative selection enables a thorough analysis of not only reconstruction accuracy but also the accuracy-complexity trade-off across diverse propagation scenarios and domain shifts. Such analysis constitutes one of the core contributions of this case study.

\subsection{Results Analysis}
The key observations drawn from the simulations, which are presented in Table \ref{tab:cross_domain_full}, can be summarized as follows:

\begin{itemize}
\item \textbf{Channel modeling heterogeneity matters more than previously acknowledged}: Models trained in one scenario typically experience noticeable performance degradation when directly tested in a different target scenario. The degradation is particularly severe when transferring between DeepMIMO-based (ray-tracing) and QuaDRiGa-based (geometry-based stochastic) scenarios. This result highlights that the mismatch between deterministic ray-tracing and stochastic channel modeling introduces significant domain shifts.

\item  \textbf{Mobility and scattering richness dominate sample size}: Although the outdoor dataset contains substantially more samples than the indoor dataset, performance gaps become larger when models are transferred to high-mobility (Highway) or rich-scattering (InF) environments. This finding indicates that propagation physics—such as Doppler effects, cluster dynamics, and material interactions—matters more than sheer data volume in determining generalization performance.

\item \textbf{Fine-tuning effectively improves target-domain reconstruction performance}: Compared with zero-shot transfer, fine-tuning the model in the target scenario leads to substantial gains in NMSE for most source-target pairs. This demonstrates that while the learned models can capture certain transferable channel structures, domain-specific adaptation remains essential to align the model with the target propagation environment. The benefit is especially pronounced when the source and target domains exhibit large physical differences.

\item \textbf{Attention-based architectures show stronger cross-physics generalization}: StarCANet achieves the best or highly competitive NMSE in most zero-shot and fine-tuning settings, particularly in transfers involving QuaDRiGa scenarios. This indicates that coordinate and spatial attention mechanisms are more effective than classical CNN or MLP designs at capturing environment-agnostic structural patterns (e.g., dominant paths and scattering geometry), rather than overfitting to simulator-specific artifacts.
\item \textbf{Attention-based models offer a favorable accuracy–memory trade-off under the evaluated metrics}: Compared with the classical CNN- and MLP-based baselines, the attention-based designs achieve competitive or superior NMSE in the majority of cross-domain settings while maintaining moderate model size and peak memory. Although their end-to-end latency does not show a clear advantage on the evaluated GPU, the overall results indicate a favorable accuracy–memory trade-off.

\end{itemize}

The observations map directly onto the LCM mechanisms standardized in Release 19. The pronounced performance degradation under zero-shot transfer across different channel modeling paradigms indicates that model identification, activation/deactivation, and fine-tuning mechanisms are highly relevant for practical deployment. The accuracy–memory profile of StarCANet further suggests that carefully designed attention models can help reduce the storage and peak-memory burden that LCM needs to manage.

\section{Future Research Directions on Practical Deployment}
\subsection{LCM}
Release 19 has established a normative framework for one-sided model LCM, including identification, activation/deactivation, performance monitoring, and basic update procedures. However, the pronounced cross-simulator domain shift observed in our experiments indicates that efficient fine-tuning and continuous performance monitoring are not optional. Current mechanisms still face multi-vendor interoperability issues, non-negligible signaling overhead, and limited support for rapid adaptation when the underlying channel statistics change.

Future work should therefore focus on: 1) lightweight model formats and parameter-efficient fine-tuning methods that reduce the cost of target-domain adaptation; 2) unified monitoring metrics that can detect domain shift early and trigger fallback or update; 3) low-overhead signaling for model switching and versioning, especially under multi-vendor and multi-scenario operation. These improvements would directly address the generalization barriers revealed by the case study and make the pre-training and fine-tuning paradigm more practical.

\subsection{Expansion of Use Cases and Complex Scenarios}
The three air-interface use cases, i.e., CSI feedback, beam management, and positioning, have received the most standardization attention. Higher-layer tasks such as intelligent scheduling, dynamic resource allocation, and random access, as well as more complex scenarios, such as multi-TRP coordination, advanced mobility, two-sided CSI compression, remain largely unexplored with AI/ML methods.

Future work should extend AI/ML applications to MAC-layer optimization. One effective approach is to reuse outputs from physical-layer models. At the same time, researchers should incorporate use-case-specific parameters, such as UE trajectories and synchronization errors. Comprehensive generalization testing should be applied to ensure cross-scenario robustness and seamless integration into future networks.


\subsection{Multi-task Learning}

Current research typically treats the core use cases separately, overlooking their clear interdependencies (e.g., positioning information can assist CSI recovery and beam prediction under high mobility). The superior cross-physics generalization of attention-based models suggests that shared structural patterns exist across tasks and environments.

Future efforts should investigate multi-task architectures with shared encoders and task-specific heads, as well as large-scale pre-training followed by targeted fine-tuning. Joint evaluation under the full set of performance-cost KPIs, rather than isolated NMSE or accuracy, will be essential for selecting models that remain practical for UE-side or network-side deployment.

\section{Conclusions}
To realize the transformative potential of AI/ML for 6G, 3GPP launched standardization efforts to integrate AI/ML into mobile networks. In this paper, we provided a comprehensive review of the standardization efforts by 3GPP on AI/ML in mobile networks, including an overview of the general AI/ML frameworks, representative use cases (i.e., CSI feedback, beam management, and positioning), and corresponding evaluations. We emphasized key research challenges related to dataset preparation, generalization evaluation and model selection. Using CSI feedback as a case study, we demonstrated the superiority of joint pre-training-fine-tuning over training from scratch. Moreover, we observed higher performance enhancements in attention-based models through fine-tuning compared with MLP-based and CNN-based models. StarCANet achieves highly competitive generalization across the tested channel modeling paradigms and scenarios, while maintaining moderate FLOPs, model size, and peak memory. These results highlight the potential of tailored coordinate attention mechanisms over standard Transformer self-attention for CSI feedback, although a more comprehensive deployment evaluation would require additional metrics such as energy consumption and system-level throughput. Finally, we outlined future research directions for the application of AI/ML in mobile networks.

\bibliographystyle{IEEEtran}
\bibliography{Myreference}

\begin{IEEEbiographynophoto}{Yuan Gao} 
is an Assistant Professor with the School of Communication and Information Engineering, Shanghai University. His research interests include integrated sensing and communication, channel foundation model and AI-native networks.\end{IEEEbiographynophoto}

\begin{IEEEbiographynophoto}{Xinyi Wu}
received the B.E. degree from the Shanghai University. She is currently pursuing the M.S. degree with the School of Communication and Information Engineering, Shanghai University, Shanghai, China. Her research interests include deep learning and wireless communication channel extrapolation.\end{IEEEbiographynophoto}

\begin{IEEEbiographynophoto}{Jun Jiang}
received the B.E. degree from the Shanghai University, Shanghai, China. He is currently pursuing the Ph.D. degree in Xi’an Jiaotong-Liverpool University. His research interests include channel foundation model and AI-native networks.\end{IEEEbiographynophoto}

\begin{IEEEbiographynophoto}{Bintao Hu} (Member, IEEE) received the Ph.D. degree in electronic and electrical engineering from the University of Sheffield, U.K. He is currently an Assistant Professor with the School of Internet of Things, Xi'an Jiaotong-Liverpool University, Suzhou, China. His research interests include wireless communications, V2X communications, mobile/vehicular edge computing, resource allocation, and intelligent optimization for IoT and 6G networks.
\end{IEEEbiographynophoto}

\begin{IEEEbiographynophoto}{Jianbo Du} (Member, IEEE) received the Ph.D. degree in communication and information systems from Xidian University, Xi'an, China, in 2018. She is currently an Associate Professor with the School of Communication and Information Engineering, Xi'an University of Posts and Telecommunications, Xi'an, China. Her research interests include mobile edge computing, blockchain, space--air--ground integrated networks, deep reinforcement learning, and resource allocation for wireless communications.
\end{IEEEbiographynophoto}

\begin{IEEEbiographynophoto}{Qiang (John) Ye} (Senior Member, IEEE) received the Ph.D. degree in electrical and computer engineering from the University of Waterloo, Waterloo, ON, Canada, in 2016. He is currently an Assistant Professor with the Department of Electrical and Software Engineering, Schulich School of Engineering, University of Calgary, Calgary, AB, Canada. His research interests include 5G/6G networking, network slicing, edge intelligence, digital-twin-assisted wireless networks, autonomous vehicular networks, and IoT.
\end{IEEEbiographynophoto}

\begin{IEEEbiographynophoto}{Shunqing Zhang} (Senior Member, IEEE) received the B.S. degree from the Department of Microelectronics, Fudan University, Shanghai, China, and the Ph.D. degree from the Department of Electronic and Computer Engineering, The Hong Kong University of Science and Technology, Hong Kong. He is currently a Professor with the School of Communication and Information Engineering, Shanghai University, Shanghai, China. His research interests include 5G/6G wireless systems, green communications, energy-efficient resource allocation, and AI-enabled wireless networks.
\end{IEEEbiographynophoto}
 
\begin{IEEEbiographynophoto}{F. Richard Yu} (Fellow, IEEE) received the Ph.D. degree in electrical engineering from The University of British Columbia, Vancouver, BC, Canada. He is currently a Professor with the School of Information Technology and the Department of Systems and Computer Engineering, Carleton University, Ottawa, ON, Canada. His research interests include intelligent and autonomous systems, connected and autonomous vehicles, IoT, security, and machine learning.
\end{IEEEbiographynophoto}

\begin{IEEEbiographynophoto}{Shugong Xu}
(Fellow, IEEE) graduated from Wuhan University, China, in 1990, and received his Master degree in Pattern Recognition and Intelligent Control from Huazhong University of Science and Technology (HUST), China, in 1993, and Ph.D. degree in EE from HUST in 1996. He is now a full professor at Xi’an Jiaotong-Liverpool University. His current research interests include machine learning, channel foundation model, AI-native networks, etc.\end{IEEEbiographynophoto}

\end{document}